\documentclass[preprint,floatfix,groupedaddress,amsmath,showpacs,showkeys,aps,prb]{revtex4}
\usepackage[final]{graphicx}
\usepackage{t1enc}
\usepackage{bm}

\begin{document}
\bibliographystyle{apsrev}

\title{\bf A computational study of the configurational and vibrational contributions to the thermodynamics of substitutional
alloys: the $\bf Ni_3Al$ case}

\author{M.F. Michelon}
\email{michelon@ifi.unicamp.br.} \affiliation{Instituto de F\'{\i}sica  ``Gleb Wataghin'', CP 6165\\
Universidade Estadual de Campinas - UNICAMP\\ 13083-970,Campinas,SP,Brazil}
\author{A. Antonelli}
\email{aantone@ifi.unicamp.br.} \affiliation{Instituto de F\'{\i}sica  ``Gleb Wataghin'', CP 6165\\
Universidade Estadual de Campinas - UNICAMP\\ 13083-970,Campinas,SP,Brazil}
\date{\today}
\begin{abstract}
We have developed a methodology to study the thermodynamics of
order-disorder transformations in n-component substitutional
alloys that combines nonequilibrium methods, which can
efficiently compute free energies, with Monte Carlo
simulations, in which configurational and vibrational degrees
of freedom are simultaneously considered on an equal footing
basis. Furthermore, by appropriately constraining the system,
we were able to compute the contributions to the vibrational
entropy due to bond proportion, atomic size mismatch, and bulk
volume effects. We have applied this methodology to calculate
configurational and vibrational contributions to the entropy of
the $\rm Ni_3Al$ alloy as functions of temperature. We found
that the bond proportion effect reduces the vibrational entropy
at the order-disorder transition, while the size mismatch and
the bond proportion effects combined do not change the
vibrational entropy at the transition. We also found that the
volume increase at the order-disorder transition causes a
vibrational entropy increase of $0.08 \; \rm{k_B/atom}$, which
is significant when compared to the configurational entropy
increase of $0.27 \; \rm{k_B/atom}$. Our calculations indicate
that the inclusion of vibrations reduces in about 30$\%$ the
order-disorder transition temperature determined solely
considering the configurational degrees of freedom.
\end{abstract}
\maketitle
\section{INTRODUCTION}
One of the goals of materials science in the field of alloys is
to predict and understand the relative stability of phases
characterized by different chemical disorder. The disorder in
an alloy can be considered as having a configurational
contribution (configurational degrees of freedom), which is the
disorder associated to the way the atoms are distributed in the
parent lattice; and a vibrational contribution (vibrational
degrees of freedom), which is the disorder associated to the
phase space region around a static lattice configuration. For a
very long time, most theoretical phase diagram calculations
were done considering only the configurational degrees of
freedom \cite{BrWi1934,Du1991}. In the 1990's, however, several
experiments measuring thermodynamical properties of alloys in
disordered metastable
states,\cite{AnOkFu1993,AnNaOkFu1994,FuAnRoNiSpMo1995,NaAnFu1995,FuAnNaNiSp1995,MuBaCh1996,NaFuRoSp1997}
demonstrated the existence of a strong interplay between
vibrational and configurational degrees of freedom. It became
clear that neglecting vibrational contributions to the
thermodynamical properties of alloys could lead to
inaccuracies, such as differences up to $30\%$ between
order-disorder (OD) transition temperatures calculated with and
without vibrational degrees of freedom. \cite{WaCe2002}
Theoretical studies of these alloys in a metastable disordered
phase were performed assuming the alloys to be either
completely ordered or totally
disordered.\cite{Ac1994,AlMoFoAsFoJo1997,RaAgBaAnFuHo1998,AlMoFoAsFoJo1998,WaCeWa1998}
The cluster variation method \cite{Kikuchi51} and its
extensions \cite{Sanchez84} have been used to calculate the
configurational contribution in partially disordered systems in
equilibrium. Different approaches have been used to incorporate
the vibrational degrees of freedom when cluster expansions are
used, such as molecular dynamics, \cite{ClMaRo1993} the coarse
graining method, \cite{WaCe2002} and the structure-inversion
approach. \cite{Connolly83,de_Fontaine94} In the last two
methods, the vibration contribution is taken into account
through the harmonic approximation and anharmonicities are
included via the quasiharmonic approximation. These two cluster
expansion methods allow a first-principles description of the
system, however, even calculations within the harmonic
approximation are still very demanding for today's computer
capabilities, and approximate approaches are still very
useful.\cite{Liu07} These recent calculations\cite{Liu07} have
shown that anharmonic effects play an important role in the
vibrational contribution to the thermodynamics of alloys. About
10 years ago, a methodology that became known as MCX was
proposed to treat simultaneously configurational and
vibrational degrees of freedom by means of Monte Carlo
simulations, which allow both atomic interchanges and atomic
displacements \cite{Purton98,Barrera00,Allan01}.

In this work, we present a methodology to investigate phase
equilibria of alloys that takes into account naturally and
simultaneously all configurational and vibrational
contributions, including all anharmonicities, through a
combination of the MCX approach and efficient tools to
determine free energies, namely, the adiabatic switching (AS)
\cite{WaRe1990} and the reversible scaling
(RS)\cite{KoAnYi1999} methods. An interesting feature of our
methodology is that it allows to study the contributions from
different vibrational mechanisms to the total vibrational
entropy. In real experiments, it is impossible to isolate the
many factors that contribute to the vibrational entropy such as
the bond proportion, the atomic size mismatch, and the bulk
volume mechanisms. On the other hand, it can be done in
computer simulations, particularly through the Monte Carlo
technique, which makes possible to isolate the constraints
associated with each mechanism. For example, the
configurational contribution to the entropy can be calculated
by allowing only the atomic interchanging dynamics. The effect
of the bonds between different atomic species can be simulated
by constraining the atoms to vibrate around their ideal
crystalline structure positions and allowing the interchanging
dynamics. The effect of the atomic size mismatch and the bond
proportion mechanisms combined can be simulated by letting the
atoms vibrate around their relaxed equilibrium positions at
fixed volume and allowing the configurational dynamics.
Finally, the volume mechanism can be simulated by allowing the
volume of the supercell to vary by imposing constant pressure
on the system, in addition to the positional and
configurational dynamics. In summary, our methodology allows
also to assess the contribution of a given vibrational
mechanism by setting the appropriate constraint in the dynamics
and then calculating the vibrational entropy difference between
the relaxed and unrelaxed system. We have applied this
methodology to quantify the vibrational entropy difference at
the thermodynamical OD transition of the $\rm Ni_3Al$ binary
alloy.

We chose the technologically important
\cite{LiSt1984a,LiSt1984b} $\rm Ni_3Al$ as the alloy model for
our study mainly because it is supposed to have the largest
vibrational entropy difference upon
disorder.\cite{AnOkFu1993,FuAnNaNiSp1995,Ac1994,AlMoFoAsFoJo1997,RaAgBaAnFuHo1998,AlMoFoAsFoJo1998,WaCe2002}
The vibrational entropy difference due to disorder at the
thermodynamical OD transition should be large enough to be
unambiguously detected, since it is, in general, a fraction of
the corresponding configurational entropy difference, which is
itself relatively small. We also chose $\rm Ni_3Al$ because it
is particularly suitable to assess the magnitude of the size
mismatch effect, since the difference between the atomic
volumes of Al and Ni is quite large \cite{MoWaCeAlFo2000}
$(V_{Al}-V_{Ni})/(V_{Al}+V_{Ni})/2=0.41$. In the case of $\rm
Ni_3Al$, most of the research in the field, both
experimental\cite{AnOkFu1993,FuAnNaNiSp1995} and theoretical,
either using the embedded atom method
\cite{AlMoFoAsFoJo1997,RaAgBaAnFuHo1998,AlMoFoAsFoJo1998} or
tight-binding \cite{Ac1994,AnMeMi2007} potentials, has found a
significant vibrational entropy difference between the totally
disordered (metastable) and the ordered phases. However, the
subject is not free of controversy, van de Walle {\it et al.}
\cite{WaCeWa1998,AnMeMi2007_exp}, using {\it ab initio}
calculations, found that the totally disordered and the ordered
phases have essentially the same vibrational entropy. Our
results indicate an increase of $0.08 \; \rm{k_B/atom}$ in the
vibrational entropy at the thermodynamical OD transition, which
is significant when compared to the corresponding
configurational entropy increase of $0.27 \; \rm{k_B/atom}$. We
have also found that the bond proportion mechanism diminishes
the vibrational entropy at the thermodynamical OD transition,
whereas the size mismatch mechanism does not change it. These
results are consistent with the local entropy calculations of
Morgan {\it et al.}\cite{MoAlFo1998,MoWaCeAlFo2000}. Regarding
the importance of the volume mechanism, theoretical studies
\cite{AlMoFoAsFoJo1997,RaAgBaAnFuHo1998,AlMoFoAsFoJo1998,Ac1994,AnMeMi2007}
have found that the volume mechanism is the main responsible
for the increase of the vibrational entropy difference between
the totally disordered and the ordered phase. This is supported
by experimental
work,\cite{GiNeCa1992,CaClMaMoRo1993,ZhZwBa1995} in which it
has been found an increase of the volume as the system becomes
totally disordered. We have found that the volume increases
1.2$\%$ at the OD thermodynamical transition. In addition, our
calculations indicate that the volume mechanism is the
responsible for the increase in the vibrational entropy
difference at the OD transition.

The paper is organized as follows. In section II we present the
general methodology. Section III describes the details of the
interatomic potential we have chosen to describe the $\rm
Ni_3Al$ alloy. In section IV, the methodology is applied to
evaluate the configurational and vibrational entropy as
function of the temperature and the contribution of each
vibrational mechanism at the OD transition. In section V we
summarize the results.
\section{METHODOLOGY}
\subsection{Dynamics and vibrational mechanisms}
In real systems, the process of chemical disordering takes
place mainly through the migration of vacancies.
\cite{OrKoPiCaPf2001,KeDiMa2003} The problem of realistically
simulating the disordering process through this mechanism is
that the average vacancy concentration is very low (less than
$10^{-5}$), \cite{MaCaAl2002} implying the requirement of very
large system sizes. For this reason, we chose the atomic
exchange dynamics to simulate the chemical disorder. This
dynamics can be implemented through the Monte Carlo method. In
this approach, the configurational degrees of freedom are
explored by selecting at random two atoms belonging to
different chemical species, their positions in the lattice are
then interchanged, the energy change upon the atomic exchange
is calculated, the Boltzmann factor associated to this change
in energy is computed, and the move is accepted or rejected
according to the Metropolis algorithm. \cite{Newman1999} In
order for the algorithm to be efficient, one should keep two
lists of atoms of each atomic species and choose randomly one
atom of each list to form the pair of atoms to be interchanged.
This can be easily done, since the number of atoms of chemical
species is kept constant. This dynamics is more efficient than
the Kawasaki dynamics and satisfies detailed balance. We will
call this dynamics as the configurational case. We consider a
Monte Carlo step (MCS) in the configurational case as $N$
attempts to exchange the atomic positions of two atoms of
different species chosen at random, where $N$ is the number of
atoms.

In order to investigate the various contributions to the
vibrational entropy we considered different dynamics related to
the various mechanisms we have mentioned earlier. To estimate
the effect of the bond proportion mechanism in the vibrational
entropy we define the following dynamics: besides the atomic
interchange dynamics, we constrain the atoms to vibrate around
their ideal crystalline structure positions. We call this
dynamics as the unrelaxed case. The vibrational dynamics for
the unrelaxed case is accomplished by choosing one atom at
random and calculating a new state by
\begin{equation} x_i^{new}=x_i^{id}+\Delta_{max}(2\xi_i-1),
\qquad i=1..3, \label{vib_dyn}
\end{equation}
where $x_i^{id}$ is the coordinate associated to the
corresponding ideal crystalline structure position, $\xi_i$ is
a random number between zero and one, and $\Delta_{max}$ is the
maximum displacement allowed, which is adjusted automatically
in such a way that $50\%$ of the trials are accepted.
\cite{AlTi1989} In this case, a MCS was considered as $N$
attempts of atomic displacements followed by $N^{\prime}$
attempts to exchange atoms. We chose $N^{\prime}=N/10$, in the
particular case of $\rm Ni_3Al$, because it is the minimum
number of attempts of atomic exchange needed for the potential
energy and the order parameter to relax to average values. The
bond proportion mechanism is claimed
\cite{MoWaCeAlFo2000,MoAlFo1998,WaCe2002} to be relevant for
changes in the vibrational entropy due to chemical disorder,
since the proportion of bonds between distinct and similar
atoms changes with the disorder. This effect could in principle
be addressed, from the theoretical point of view, through
calculations based on a spring
model,\cite{MoWaCeAlFo2000,MoAlFo1998,WaCe2002} which
presumably assigns a softer and a stiffer character to the
bonds between like and unlike atoms,
respectively.\cite{Ni3AlMoetal}

We now set a different type of dynamics by letting the atoms
relax around their thermal equilibrium position as
\begin{equation}
x_i^{new}=x_i^{old}+\Delta_{max}(2\xi_i-1), \qquad i=1..3.
\label{rel_dyn}
\end{equation}
This kind of dynamics allows the atoms to vibrate around their
thermal equilibrium positions, taking into account naturally
their accommodation due to their different atomic sizes. This
is another mechanism that is claimed to influence the
vibrational entropy and is called the size mismatch mechanism.
It can be understood by the following picture:\cite{WaCe2002}
when two large atoms are constrained in a small space they can
experience a compressive stress, increasing the stiffness of
the bond, and reducing the allowed region to vibrate; on the
other hand, when two small atoms are constrained in the same
space they can experience a tensile stress, increasing the
softness of the bond, and increasing the allowed region to
vibrate. The interplay between the compressive and tensile
stresses will dictate the final vibrational effect. We call
this relaxation of the configurational and positional
constraints as the partially relaxed case. With this dynamics
we simulate the combined effect of bond proportion and size
mismatch in the vibrational entropy. In this case also, a MCS
was considered to be the same as in the unrelaxed case.

Finally, we simulate the contribution from the volume mechanism
by allowing the volume change in a constant pressure
simulation. This effect is explained \cite{WaCe2002} by the
following reasoning: as the average interatomic distances
increase, the bonds become softer and the atoms have more space
to visit, that causes an increase in the vibrational entropy.
We call this relaxation of the configurational, positional, and
volumetric constraints as the fully relaxed case. Through this
dynamics we simulate the combined effect of the bond
proportion, size mismatch, and volume mechanisms in the
vibrational entropy. The volumetric dynamics is accomplished by
rescaling the atomic positions according to the volume change
calculated as
\begin{equation}
L^{new}=L^{old}+\Delta_{max}^{L}(2\xi-1),
\label{vol_dyn}
\end{equation}
where $L$ is the length of the simulation box, and
$\Delta_{max}^{L}$ is the maximum allowed change in that
length, which is adjusted similarly to the unrelaxed and
partially relaxed cases at typically 10 MCS. In the fully
relaxed case, a MCS was defined as $N$ positional trials
followed by $N^{\prime}=N/10$ exchanging trials, and one
volumetric trial.\cite{AlTi1989}

\subsection{Free energy calculations}
The thermodynamical quantity that underlies all this work is
the free energy, which is calculated through the AS
\cite{WaRe1990} and RS \cite{KoAnYi1999} methods. The AS method
allows one to calculate the free energy by computing the work
done by adiabatically switching the Hamiltonian of the system
of interest to the Hamiltonian of a reference system (or
vice-versa) at a single given temperature. On the other hand,
the RS method allows one to evaluate the free energy in a range
of temperatures provided that it is known at a single given
temperature. These methods are very efficient since they
evaluate the free energy from only one simulation run, whose
length is determined by the required accuracy. In contrast with
other methods, such as the harmonic,\cite{MaMoWeIp1971} or the
quasi-harmonic
\cite{AlMoFoAsFoJo1997,AlMoFoAsFoJo1998,WaCeWa1998}
approximations, the AS and RS, take into account naturally all
anharmonic effects, which are crucial for the calculation of
vibrational entropy differences.

The AS method is based on the well known thermodynamic
integration method (TI).\cite{FrSm2001} In the TI method, the
absolute free energy of a system of interest can be estimated
by computing the work done to transform the Hamiltonian of a
reference system, of which one knows the free energy, into that
of the system of interest. This can be achieved by considering
the artificial Hamiltonian, $H(\lambda) = \lambda H_{sys} +
(1-\lambda) H_{ref}$, where $H_{sys}$ is the Hamiltonian of the
system of interest, $H_{ref}$ is the Hamiltonian of the
reference system, and $\lambda$ is a dimensionless coupling
parameter. By varying $\lambda$ from 0 to 1, one can transform
one Hamiltonian into the other one. The work performed to
switch between the two systems is given by the integral
\begin{equation}
F-F_{ref}=\int_{0}^{1}d\lambda \left\langle\frac{\partial H}{\partial \lambda}\right\rangle_{\lambda}.
\end{equation}
If now $\lambda$ is considered to be a function of time, and
its value continuously varied from 0 to 1 during the time of
simulation $t_{s}$, the free energy difference between the two
systems is given by
\begin{equation}
F-F_{ref}=\int_{0}^{t_s}dt\frac{d \lambda}{dt}(U_{sys}-U_{ref})=W_{irr}=W_{rev}+E_{diss},
\label{FAS4}
\end{equation}
where $U_{sys}$ is the potential energy of the system of
interest, $U_{ref}$ is the potential energy of the reference
system, $W_{irr}$ and $W_{rev}$ are the irreversible and
reversible work, respectively, and $E_{diss}$ is the energy
dissipation. Time in Eq.(\ref{FAS4}) can be regarded as the
actual time, as in a molecular dynamics simulation, or the
fictitious time created by the successive steps in a Monte
Carlo simulation. The potential energy difference between the
reference system and the system of interest appears in
Eq.(\ref{FAS4}), instead of the Hamiltonian difference, because
we consider the kinetic degrees of freedom to be in
equilibrium, and, therefore, the kinetic energy terms cancel
each other. The energy dissipation is one source of error,
characteristic of nonequilibrium dynamic processes, and can be
estimated \cite{KoCaAnYi2000} by performing the direct and
inverse transformations between the two systems:
\begin{equation}
E_{diss}=\frac{W_{irr}^{ref \rightarrow sys}+W_{irr}^{sys \rightarrow ref}}{2}.
\label{FAS8}
\end{equation}
In all AS and RS calculations we adopted this criterion to
quantify the free energy error, which can be reduced by
increasing the simulation time. Another source of error
\cite{KoCaAnYi2000} are the statistical fluctuations of the
quantities in the integrand of Eq.(\ref{FAS4}), which can be
handled by simulating other trajectories and averaging the
results.

There are subtleties in the AS method we must be aware of in
order to obtain the correct free energy. The external
conditions, like temperature, and the parameters of the
reference system must be set in a way that the coupled system,
described by $H(\lambda)$, does not undergo a phase transition
along the transformation path. (We took particular care about
the choice of the reference temperature for the high
temperature disordered phases in the unrelaxed, partially
relaxed, and fully relaxed cases to avoid mechanical melting.)

Now let us discuss briefly the RS method and its
application.\cite{KoAnYi1999} In contrast to the AS, the RS
method allows the calculation of the free energy in a range of
temperatures. This can be accomplished by realizing that the
free energy of the scaled system at a temperature $T_{0}$,
whose potential energy is given by $U_{scaled}=\lambda U_{sys}$
(in the case of RS $\lambda$ is not restricted to the interval
$[0,1]$), is related to the free energy of our system of
interest at a temperature
$T=T_{0}/\lambda$.\cite{KoAnYi1999,KoCaAnYi2000} The free
energy of the scaled system at a given value of $\lambda$ can
be readily determined by computing the work performed to change
$\lambda$ from 1 to $\lambda = T_{0}/T$, as it is done in the
AS method (provided that the free energy is known for
$\lambda=1$). It is shown in Ref. \onlinecite{KoAnYi1999} that
the free energy of a system at temperature $T$ can be estimated
from the irreversible work $W_{irr}(t)$ done to bring the
system from $T_0$ to $T(t)$, as
\begin{equation}
\frac{F(T(t))}{T(t)}= \frac{F(T_0)}{T_0}+\frac{W_{irr}(t)}{T_0}-\frac{3}{2}k_B N \ln\frac{T(t)}{T_0},
\label{FRS3}
\end{equation}
where $T(t)=T_0/\lambda(t)$, and $F(T_0)$ is the known free
energy reference. The logarithmic term of Eq.(\ref{FRS3})
corresponds to the contribution of the kinetic degrees of
freedom and must be omitted when only the configurational
changes are considered. The estimation of the energy
dissipation at a given temperature is calculated, as in the AS
method, using Eq.(\ref{FAS8}). In the case where the external
pressure is set zero, as in the calculation of the free
energies for the fully relaxed case, the Gibbs free energy
formula reduces to Eq.(\ref{FRS3}).

\section{THE $\rm \bf Ni_3Al$ SYSTEM}
\subsection{The choice of potential}
Some well known and often used interatomic potentials for
modeling $\rm Ni_3Al$ \cite{ClRo1993,PaChEvPa2003} are not
suitable to describe the configurational degrees of freedom of
this alloy.\cite{MiAn2007} The reason for that is that this
potential, in both
parameterizations,\cite{ClRo1993,PaChEvPa2003} does not yield
the $L1_2$ phase as the ground state phase, giving rise to
nonphysical thermodynamical phases at low
temperatures.\cite{MiAn2007} In order to modeling appropriately
the $\rm Ni_3Al$ system we looked for a potential which
provides not only the correct ground state, but describes, at
least qualitatively, the thermodynamics of the OD and
vibrational phenomena. We chose the tight-binding
Finnis-Sinclair \cite{FiSi1984} potential whose
parameterization was obtained by Vitek {\it et
al.}\cite{ViAcCs1991} Among the thermal features of this
potential we may cite the linear thermal expansion coefficient
(at 1050 K) of $21.7 \times 10^{-6} \rm{K}^{-1}$, which agrees
well with the experimental value of $19 \times 10^{-6}
\rm{K}^{-1}$; \cite{MoSuSaNa1989} the equilibrium lattice
parameter (at 1000 K) of $a_0=3.6096$ \AA, which is in good
agreement with the experimental value of $a_0=3.6120$
\AA;\cite{MoSuSaNa1989} and the calculated mechanical melting
temperature $T_{m}^{mech}=1600\pm25$ K at the
Lindemanns'$\delta$ function value of $0.12$. We have
determined the thermodynamical melting temperature for this
model of $\rm Ni_{3}Al$ to be $T_m=1328\pm6$ K. The
thermodynamical melting point of a substance is obtained by
determining at which temperature the solid and liquid phases
have the same free energy. It is important to point out that
the thermodynamical melting temperature we have obtained for
the model is 20$\%$ lower than the experimental value
$T_m^{exp}=1636$ K. \cite{BrBeWe1988} This discrepancy in the
melting transition temperature is not surprising since the
potential parameters are fitted from a database which does not
include data from the liquid phase. We will return to this
point later, after we present the results for the OD
transition. However, the important conclusion we should advance
at this point is that, despite numerical discrepancies, the
results from our simulations for the OD transition temperature
and the melting point are qualitatively consistent with
experimental findings.

\subsection{Order parameters}
In the case of $L1_2$ alloys, the order parameters can be
defined as follows. The long-range order parameter is
constructed from the $L1_2$ phase by labeling the sublattice
associated to the Al (Ni) atom as an $\alpha$ ($\beta$)
sublattice. The long-range order is then measured by the
formula, first introduced by Bragg and Williams,
\cite{BrWi1934}
\begin{equation}
\eta=\frac{p_{\alpha}-0.25}{1-0.25},
\label{OPeta}
\end{equation}
where $p_{\alpha}$ means the fraction of Al atoms in the
$\alpha$ sublattice. In this way, $\eta=1$ for the ordered
$L1_2$ phase and $\eta=0$ for the disordered phase. This order
parameter is very useful to quantify the long-range order,
however one must be careful with its interpretation. First,
when one performs computer simulations to explore the
configurational degrees of freedom through cooling experiments,
at a relative low rate, starting from the disordered phase at
high temperatures, the system should always end up in the
$L1_2$ phase, but the Al atoms not always are found in the
arbitrarily defined $\alpha$ sublattice. In other words, one
does not know, in advance, which one of the four possible
sublattices will be $\alpha$ sublattice. Hence, in this kind of
experiment one must measure the long-range order parameter in
the four possible sublattices. Second, when the system is in an
antiphase boundary (APB)-like configuration \cite{MiAn2007} a
large fraction of Al atoms may be in a ordered block at sites
of a $\beta$ sublattice, giving low and even negative values
for $\eta$. Therefore, $\eta =0$ does not distinguish between a
totally disordered and a particular APB configuration. In Fig.
\ref{4troca}d, we show the results for the long-range order
parameter for the $\alpha$ and $\beta$ sublattices.

Concerning the short-range correlations we measure the
short-range order parameter, first introduced by Bethe and
Wills,\cite{BeWi1935} as
\begin{equation}
\sigma=\frac{p_{Al-Ni}-9}{12-9},
\label{OPsigma}
\end{equation}
where $p_{Al-Ni}$ means the average number of unlike bonds
between an Al atom and its first-neighbors. Note that
$\sigma=0$ implies $\eta=0$, however, $\eta=0$ can correspond
to $\sigma = 1$ as in a particular APB configuration.

\subsection{Implementation details}
We have performed tests of our computational code by
calculating the melting temperature of the Ni system using the
Cleri and Rosato potential, \cite{ClRo1993} which agrees
exactly with the value reported in Ref.
\onlinecite{BeKaHe2006}. Furthermore we compared our
Finnis-Sinclair calculations of antiphase-boundary and stacking
fault energies with Vitek {\it et al.} \cite{ViAcCs1991}
results, and verified an exact agreement.

The reference system chosen for the calculation of free energy
references in the solid phases was the Einstein
crystal.\cite{KoAn1996,KoAn1997,FrSm2001} The chosen values for
the vibration angular frequencies are $w_{Al}=75.4 $ rad THz,
and $w_{Ni}=31.4$ rad THz, for the Al atom and for the Ni atom
respectively. These are the frequencies of the main vibrational
modes of the elements, estimated from the experimental phonon
density of states from Ref. \onlinecite{StKaLoAr1981}, which
are expected to be optimal to mimic the vibration of the atoms
in the alloy. The reference system chosen for the calculation
of the free energy reference of the liquid phase (used to
estimate the melting temperature) was the $r^{-12}$ repulsive
fluid.\cite{HoGrJo1971,YoRo1984} The repulsive fluid parameters
are chosen in such a way that the position and height of the
first peak of the radial distribution function of the $r^{-12}$
repulsive fluid potential coincide with those of the
Finnis-Sinclair potential. This choice of parameters enhances
the probability of the reference system to be within the
borders of the phase diagram of the system of interest, thus
minimizing the risk of encountering a phase
transition.\cite{HoGrJo1971,KaRuHe2004,BeKaHe2006}

In the AS and RS calculations we chose time simulations such
that the energy dissipation was less than $10^{-4}\;
\rm{eV/atom}$, which typically leads to simulation lengths of
$2 \times 10^5$ MCS. The functional form of $\lambda (t)$ was
always chosen to be a linear interpolation between the initial
and final simulation times, which correspond to the initial and
final temperatures. To circumvent surface effects we applied
periodic boundary conditions and the minimal image convention.
\cite{AlTi1989} Since both, the Einstein crystal and the
$r^{-12}$ repulsive fluid do not have any cohesion, the
simulations involving these systems have to be performed at
fixed volume, which is chosen to be the average volume of a NPT
equilibrium simulation at the given pressure and temperature of
interest. Aside from the systematic errors due to dissipation,
statistical errors in free energy calculations were handled by
taking averages over typically 10 samples. The error bars in
the entropy differences were obtained from the fluctuation of
entropy data below and above the transition in the standard
way. In most of the results that will be presented in the
following section, a simple running average smoothing procedure
was used in order to remove the unwanted fluctuations
introduced by the numerical derivative calculations. All the
calculations were performed using a cubic simulation cell
containing 500 atoms. Finally, we tested a larger system size
using a 1372-atom simulation cell and found no significant
finite-size effects in entropy differences and transition
temperatures.

\section{RESULTS AND DISCUSSION}
\subsection{The configurational case}

\begin{figure}[htbp]
  \includegraphics[scale=0.55]{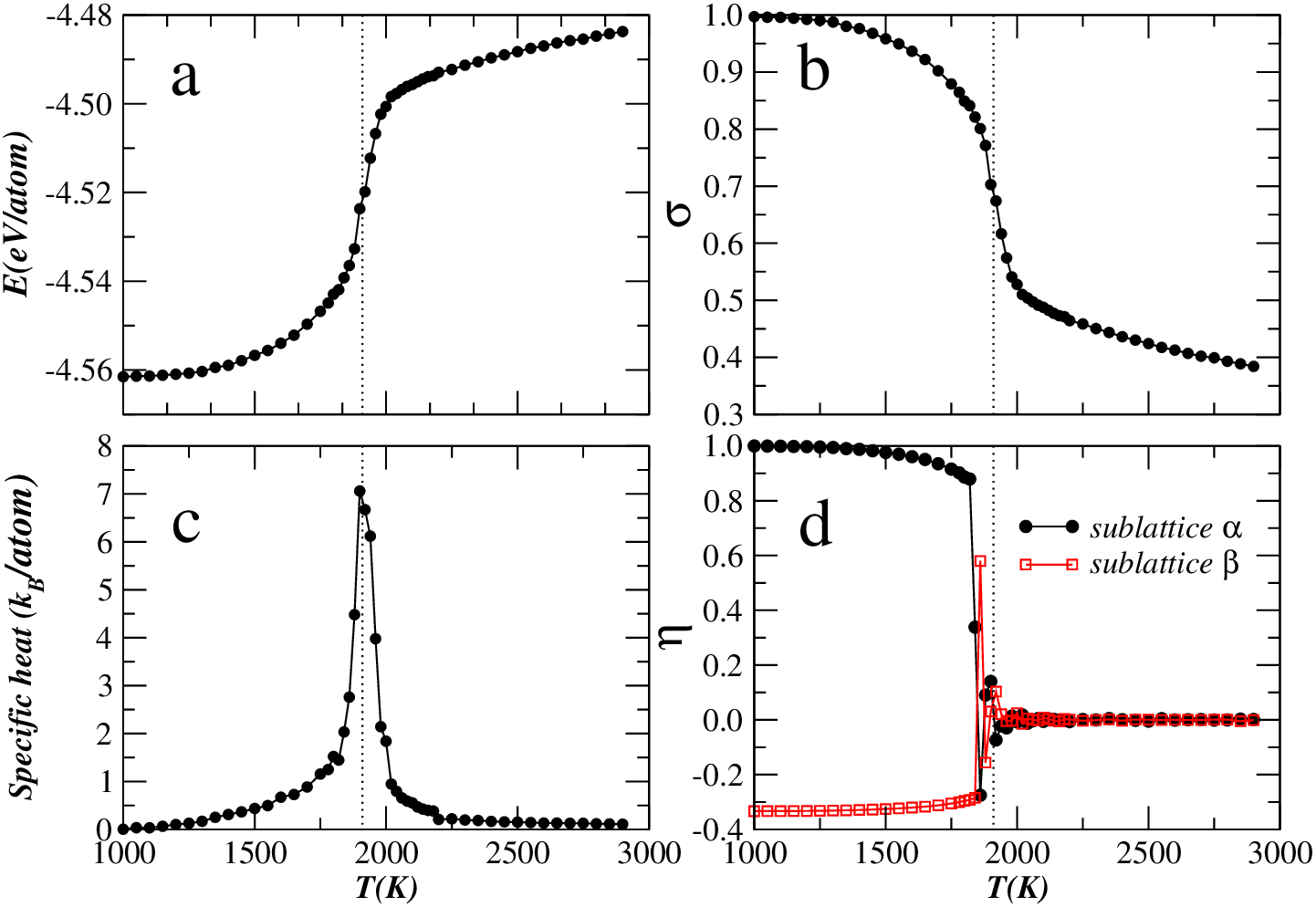}
  \caption{\label{4troca} (Color online) Thermal equilibrium quantities for the
  configurational case. The dotted lines indicate the OD temperature
   calculated from free energy calculations(Fig.\ref{troca_F_S}). The order parameter $\eta$
   is shown for two of the four sublattices of the $L1_2$ structure, the order
   parameters not shown exhibit an identical behavior to that of sublattice $\beta$.}

\end{figure}

Let us discuss briefly the equilibrium numerical experiments
performed in order to bracket the OD temperature for the
configurational case. We set the $L1_2$ phase at a fixed volume
corresponding to the equilibrium volume obtained at zero
pressure and $T_{0}=10^3$~K. Then we turned on the exchange
dynamics and performed a series of equilibrium simulations on a
relatively fine grid over a temperature interval of 2000 K. The
measured thermodynamical quantities are shown in Fig.
\ref{4troca}. The abrupt changes in the behavior of the
potential energy, specific heat, and order parameters indicate
the OD transition at $T_{od}^{conf}=1925\pm30$~K. Note the
abrupt change of the long-range order parameter $\eta$, and the
nonzero value of $\sigma$ after the transition. In order to
estimate the effect of the chosen fixed volume on our results,
we performed an analogous series of calculations at the
equilibrium volume at 0 K, which is substantially smaller than
the one at zero pressure and $T_{0}=10^3$~K. We found the OD
transition to be only 5\% larger than the previous one, and
such small difference indicates that the chosen value for the
volume is not relevant for our conclusions.

\begin{figure}[htbp]
  \includegraphics[scale=0.55]{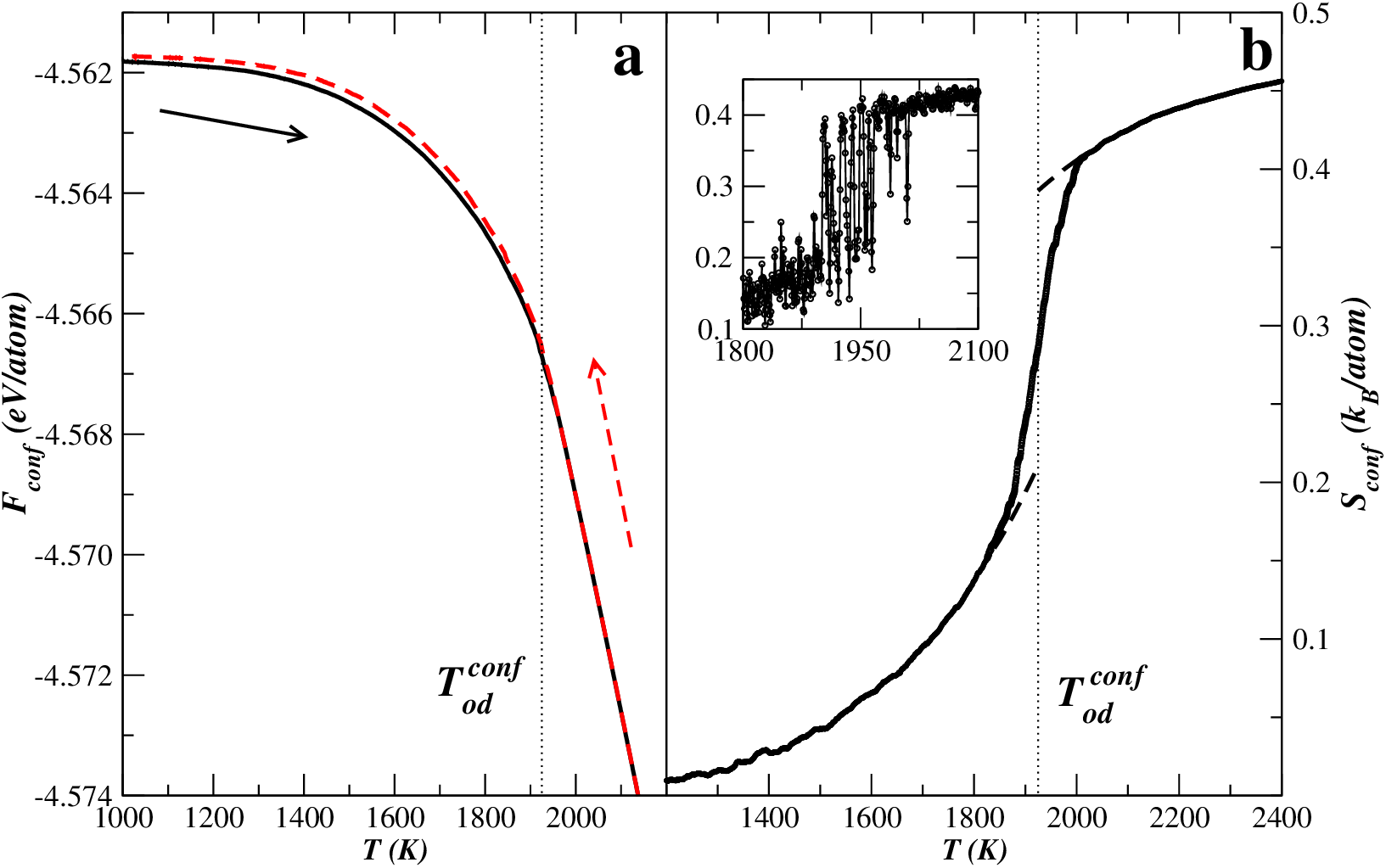}
  \caption{\label{troca_F_S} (Color online) Helmholtz free energy and entropy versus
   temperature for the configurational case. In (a) the solid line and the dashed line correspond
    to single realizations of the quasi-static heating and cooling processes, respectively.
    In (b) solid line depicts the entropy obtained from smoothing the $-d\langle F_{conf}\rangle /dT$ data,
    dashed lines correspond to the coexistence region, and the OD
    transition temperature $T_{od}^{conf}=1925\pm30 \rm{K}$ is estimated from the center of the coexistence region.
    The inset shows the entropy of a typical single realization where the coexistence behavior
    can be observed.}
\end{figure}

We now discuss the free energy calculations. We consider the
free energy reference in this case to be at infinite
temperature, because in this limit the configurational entropy
per atom of a system with N atoms has an analytical expression
corresponding to the ideal solid solution, which is
\begin{equation}
S_{conf}(\infty)={\frac{1}{N}}k_B{\rm ln}\frac{N!}{N_{Al}!N_{Ni}!}.
\label{Eq_Sconf}
\end{equation}
This quantity measures the number of distinct configurations
obtained by arranging $N_{Al}$ and $N_{Ni}$ atoms in the
lattice. Thus, for a system containing 500 atoms, we have
approximately $S_{conf}(\infty)=0.556 \; \rm k_B/atom$. The
advantage of using the RS method in this case results from the
fact that the method maps the problem of determining the free
energy for an infinite interval of temperature onto a problem
of finding the free energy for a finite interval of the scaling
parameter $\lambda$. We determine the work done to take the
system from $T_0=10^3$~K ($\lambda = 1$) to the virtually
infinite temperature ($\lambda = 0$). Combining this work and
the entropy from Eq.(\ref{Eq_Sconf}), we are able to calculate
the free energy at $T_0$ using Eq.(\ref{FRS3}), noting that in
the configurational case the logarithmic term should be
dropped, since in this case the atoms are not allowed to
vibrate. From the free energy reference at $T_0$ and by
computing the work done to take the system from $\lambda = 1$
to any $\lambda < 1$, we are able to calculate F as a function
of T as shown in Fig. \ref{troca_F_S}.

\begin{figure}[htbp]
  \includegraphics[scale=0.55]{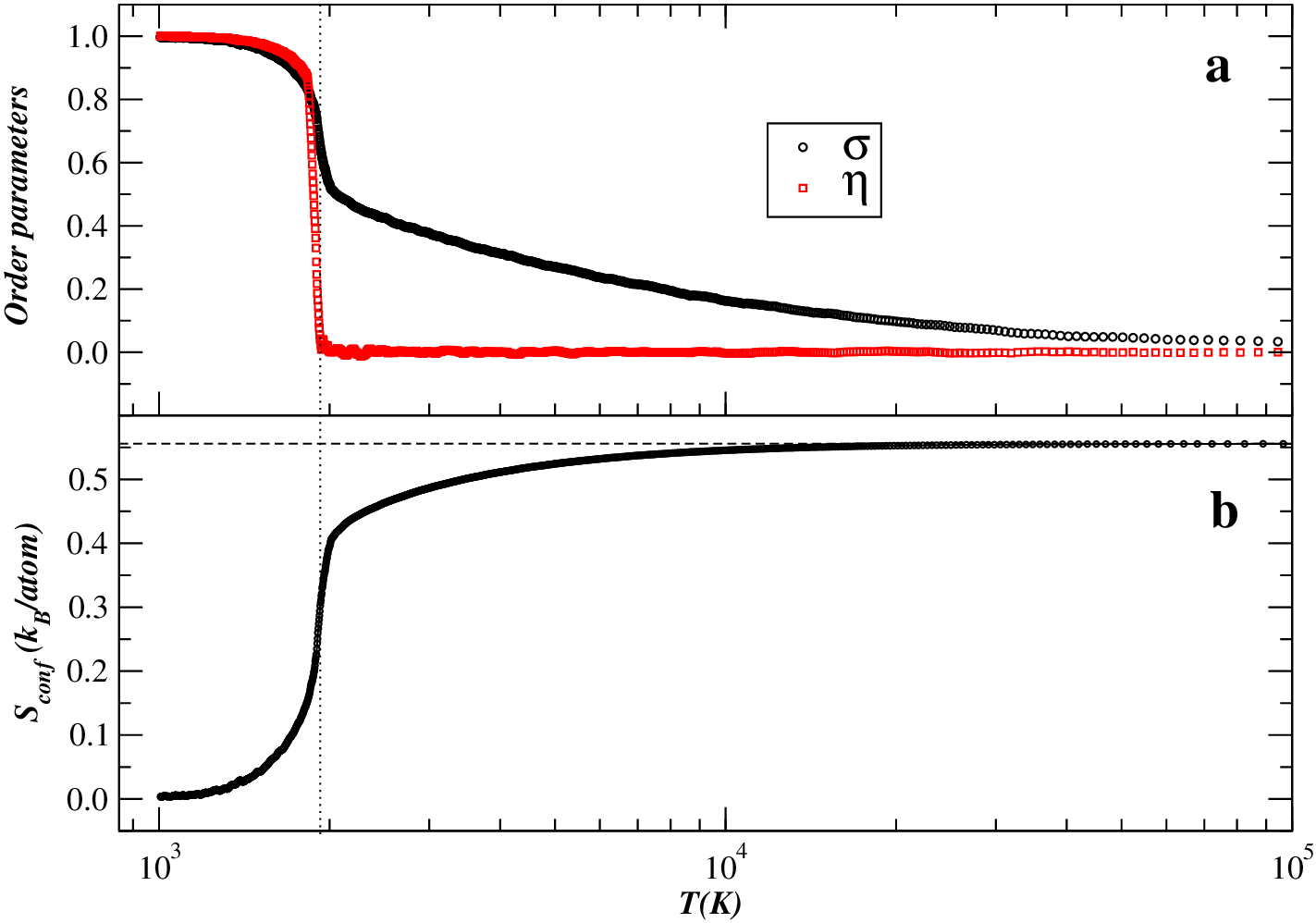}
  \caption{\label{sigma_eta_S} (Color online) Behavior of the order parameters and entropy as functions
    of temperature in a logarithmic scale for the configurational case. In (a) the
    long-range order parameter $\eta$ vanishes at the OD transition, in contrast
    to the short-range order parameter $\sigma$, which vanishes only at very high temperatures.
    This behavior is reflected in the configurational entropy showed in (b). The dashed line in (b)
    indicates the value of $S_{conf}(\infty)$.}
\end{figure}

In order to estimate the energy dissipation we compute the work
performed to take the system from $\lambda = 0$ (infinite
temperature) to $\lambda = 1$ ($T_0$). As we can see in Fig.
\ref{troca_F_S}a the energy dissipation in the direct and
reverse processes is less than $10^{-4}\;\rm{eV/atom}$. The
configurational entropy, shown in Fig. \ref{troca_F_S}b, is
obtained by computing numerically $- d \langle F_{conf}\rangle
/d T$, where the brackets denote an average over uncorrelated
samples of the configurational free energy. This averaging
procedure is done in order to reduce the statistical noise in
the numerical derivative (subsequently the remaining noise is
further reduced by applying a running average smoothing
procedure). The OD transition temperature, which in this case
is considered to be at the center of the coexistence region of
the two phases, was found to be $T_{od}^{conf}=1925\pm30$~K,
which agrees with the transition temperature determined by
analyzing the behavior of thermodynamical quantities in Fig.
\ref{4troca}. The configurational entropy difference calculated
at the OD transition is $\Delta S^{conf}_{conf}=0.18\pm
0.01\;\rm{k_B/atom}$. In Fig. \ref{sigma_eta_S}a we show the
behavior of the order parameters as function of the
temperature. We can see that the long-range order parameter
goes to zero at the OD transition temperature, whereas the
short-range remains finite above the transition, approaching
zero only for extremely high temperatures. Due to the
persistence of the short-range order, we note that the
configurational entropy reaches its maximum only at very high
temperatures. The similar behavior of the configurational
entropy and short-range order parameter with temperature allows
us to establish a direct relationship between these two
magnitudes. This relationship will be used in the calculation
of free energy references for the other cases we will study.
This result may also be especially useful at much higher
pressures where the OD temperature is much lower than the
melting point.\cite{GeChSl2005}

\subsection{Unrelaxed, partially relaxed, and fully relaxed cases}
\begin{figure}[htbp]
  \includegraphics[scale=0.55]{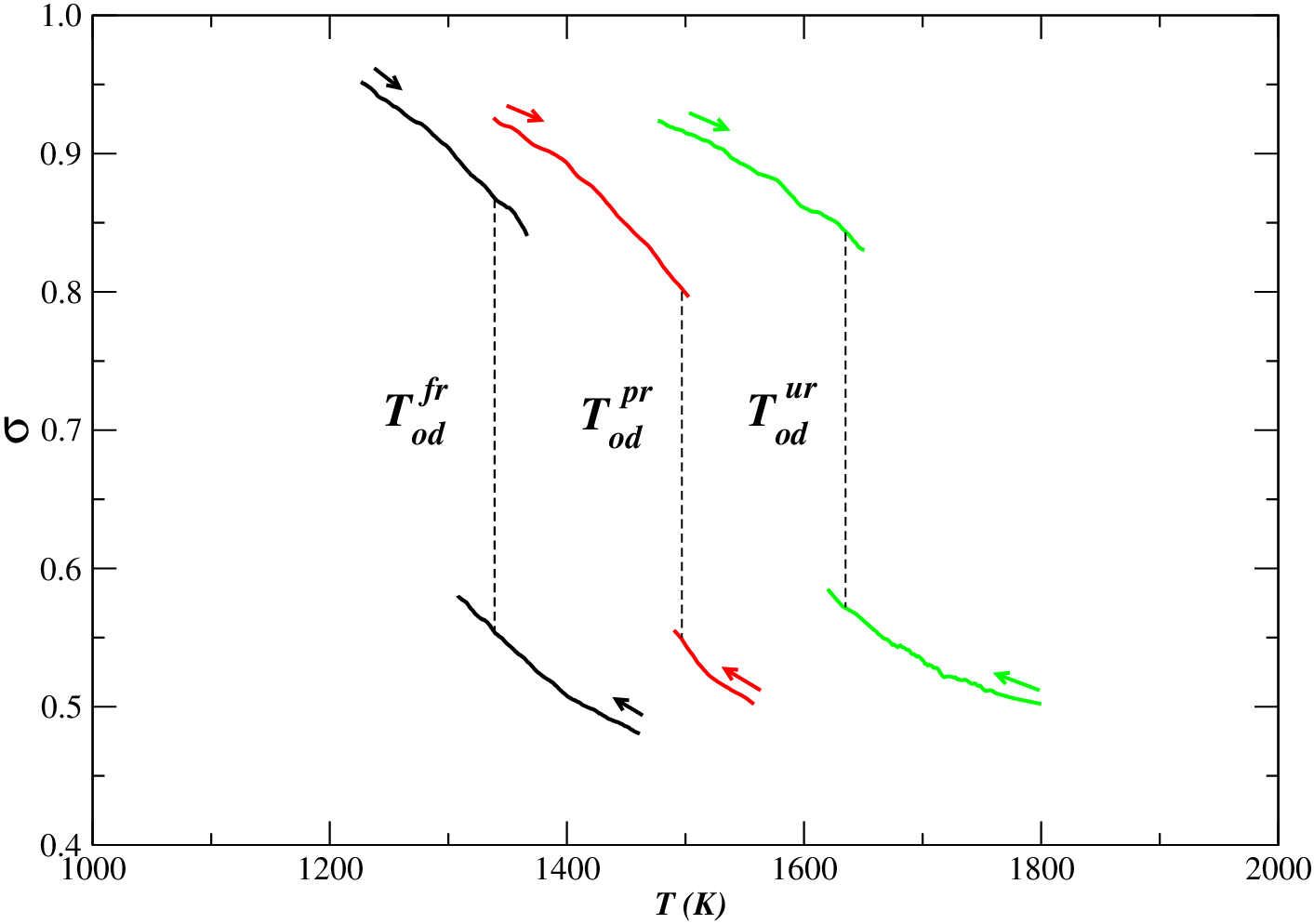}
  \caption{\label{Heat_dT0_1K_MCS_sigma} (Color online) Behavior of the short-range order
   parameter $\sigma$ as a function of temperature in heating and cooling numerical experiments
   at a rate of 0.02 K/MCS. From left to right, the three pairs of curves are for the fully relaxed,
   partially relaxed, and unrelaxed cases. The dashed lines indicate
   the OD transition temperatures, obtained from the crossing of the free energy curves,
   which are $T_{od}^{fr}=1339\pm20\; \rm{K}$, $T_{od}^{pr}=1497\pm40\; \rm{K}$, and $T_{od}^{ur}=1635\pm60\; \rm{K}$.
   The curves are the smoothed data from averages over 10 samples.}
\end{figure}

Now we turn to the other cases, where the constraints on the
degrees of freedom are gradually lifted. When vibrations are
allowed, the limit of infinite temperature is no longer a
suitable reference for the free energy, since the system would
not remain a solid. In order to bracket the OD transition
temperatures for each case, we performed a series of heating
and cooling experiments. In Fig. \ref{Heat_dT0_1K_MCS_sigma} we
plot the short-range order parameter $\sigma$ as a function of
the temperature, for the unrelaxed, partially relaxed, and
fully relaxed cases. The metastability exhibited in each case
allows suitable choices of reference temperatures for the
calculation of reference free energies using the AS method. For
each case, two reference temperatures are chosen, one below and
one above the guessed OD transition temperature provided by the
metastability region. These free energy references are
subsequently used in the RS method to calculate the free energy
curves starting from both reference temperatures. Starting at
the lower reference temperature, the RS method generates a free
energy curve for increasing temperatures; from the higher
reference temperature, on the other hand, the RS method
provides a free energy curve for decreasing temperatures. The
crossing of these two curves gives the OD transition
temperature. These OD transition temperatures are indicated by
dotted lines in Fig. \ref{Heat_dT0_1K_MCS_sigma}. The error
bars for the transition temperatures were obtained from the
free energy reference error bars in the same way as in Ref.
\onlinecite{KoAnYi2001}: the RS method is performed again, this
time starting from the extremes of the error bar of free energy
reference given by the AS method. The temperature error bar is
then obtained by the two crossings points that are the farthest
from each other among the four curves intersections around the
transition. Next, we discuss the details of these calculations
and further results.

The total free energy at the reference temperature $T_{ref}$
for the unrelaxed, partially relaxed, and fully relaxed cases
is calculated by adding the contributions from the vibrational
and configurational degrees of freedom as
\begin{equation}
F(T_{ref})=F_{vib}(T_{ref})-T_{ref}S_{conf}(\sigma (T_{ref})),
\label{Eq_Fconf}
\end{equation}
where $F_{vib}(T_{ref})$ is the vibrational free energy
calculated through the AS method to a reference system which
does not take into account the configurational entropy, e.g.,
the Einstein Crystal, and $S_{conf}$ is the configurational
entropy corresponding to the short-range order parameter at
$T_{ref}$. The mapping between $S_{conf}$ and $\sigma$ is
obtained from their temperature dependence in the
configurational case, using the data given in Fig.
\ref{sigma_eta_S}.

\begin{figure}[htbp]
  \includegraphics[scale=0.55]{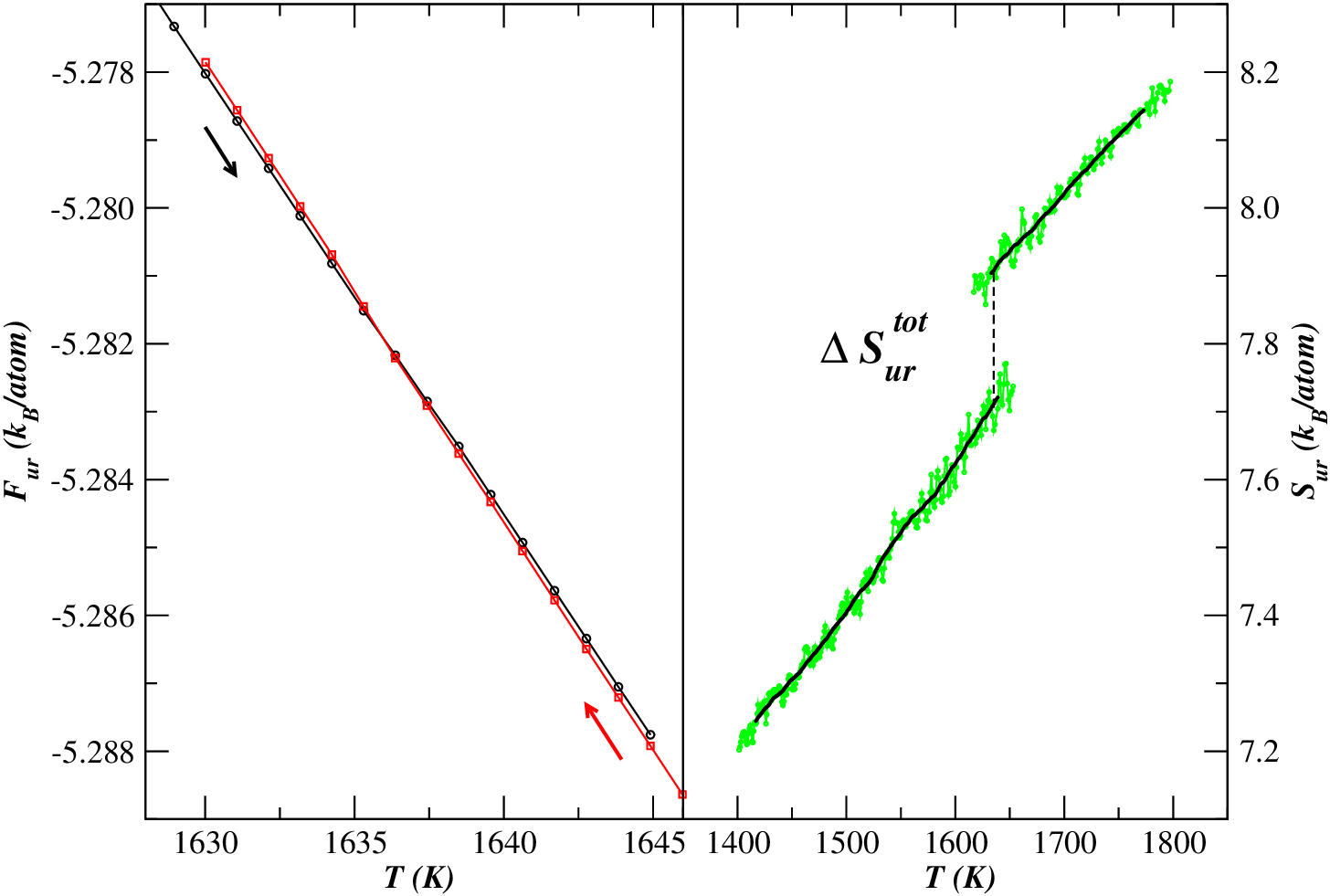}
  \caption{\label{F_S_vib}(Color online) Free energy and entropy as functions of temperature
   for the unrelaxed case. The dashed line indicates the OD temperature of
   $T_{od}^{ur}=1635 \pm 60\; \rm{K}$ obtained from the crossing of the free energy curves. The lines in the entropy plot
   are obtained from the smoothing of the $-d\langle F_{ur}\rangle /dT$ data.}
\end{figure}

In order to evaluate the contribution of the bond proportion
mechanism to the vibrational entropy, we calculated the free
energy of the system in the unrelaxed case. In Fig.
\ref{F_S_vib} we show the free energy and entropy curves below
and above the OD transition. The crossing of the two free
energy curves determines the OD transition. The entropy is then
computed by the numerical derivative of the free energy curves.
The OD transition temperature obtained where the two curves
cross is $T_{od}^{ur}=1635\pm60~\rm{K}$. Once the transition
temperature is obtained, one can go back to the data in Fig.
\ref{Heat_dT0_1K_MCS_sigma} to determine the short-range order
parameter for the ordered and disordered phases at the OD
transition, and from that the configurational entropy
difference at the OD transition. The total entropy difference
and the configurational entropy difference at the transition
are $\Delta S_{ur}^{tot}=0.20\pm0.02~\rm{k_B/atom}$ and $\Delta
S_{ur}^{conf}=0.27\pm0.01 \rm{k_B/atom}$, respectively. So the
entropy difference due to only the vibration
\begin{equation}
\Delta S_{ur}^{vib}=\Delta S_{ur}^{tot}-\Delta S_{ur}^{conf},
\label{delta_Svib}
\end{equation}
is $-0.07\pm0.02 \rm{k_B/atom}$. This negative value is
consistent with the local entropy calculations of Morgan {\it
et al.}, \cite{MoAlFo1998,MoWaCeAlFo2000} who concluded that
the increase of Al nearest-neighbors decreases the local
vibrational entropy of Al or Ni atoms.

The combined effect of size mismatch and the bond proportion
mechanisms was studied by simulating the system in the
partially relaxed case. The OD temperature obtained by the
crossing of the free energy curves is
$T_{od}^{pr}=1497\pm40\;\rm{K}$. The total partially relaxed
entropy difference and the configurational entropy difference
at the transition are $\Delta S_{pr}^{tot}=0.23\pm0.02 \;
\rm{k_B/atom}$ and $\Delta S_{pr}^{conf}=0.22\pm0.01 \;
\rm{k_B/atom}$, respectively. So the entropy difference due to
only the relaxation of the atoms at fixed volume is
approximately $0.01\pm0.02 \; \rm{k_B/atom}$. This result is
also consistent with Morgan {\it et al.},
\cite{MoAlFo1998,MoWaCeAlFo2000} who observed that the local
vibrational entropy ``seems very flat, or even slightly
increasing'' as the number of relaxed Al first-neighbors
increases.

Simulations of the system in the fully relaxed case were
employed to evaluate the combined effect of the volumetric
relaxation of the supercell with the size mismatch and the bond
proportion mechanisms. The calculated OD temperature
$T_{od}^{fr}=1339\pm20\; \rm{K}$ essentially coincides, within
the error bars, with the melting temperature of $T_m=1328\pm6$
K. This is in agreement with experimental
findings.\cite{Cahn1987,BrBeWe1988} Although the
Finnis-Sinclair potential cannot reproduce quantitatively the
values obtained experimentally, it provides results that are
consistent with the experimental results.

\begin{figure}[htbp]
  \includegraphics[scale=0.55]{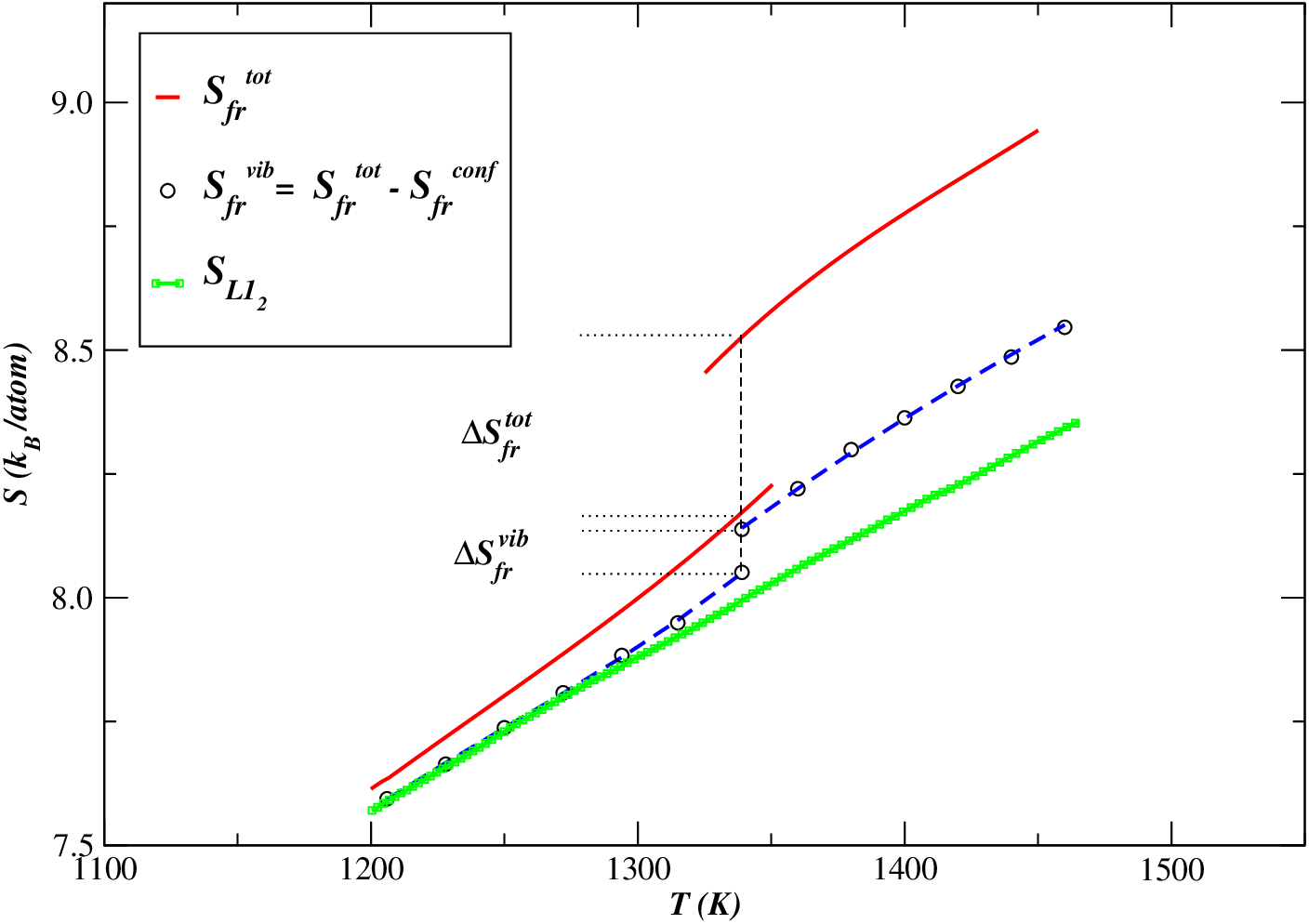}
   \caption{\label{S_fr_conf}(Color online) Entropy as a function of temperature for the fully relaxed case (solid curve) and
   perfectly ordered $L1_2$ phase (squares). The open circles depict the fully relaxed vibrational entropy, which is the difference between
the fully relaxed total entropy and the fully relaxed configurational entropy.
The vertical short dashed line indicates the fully relaxed OD temperature. The open circles represent the smoothed data from
averages over 10 samples. The solid and long dashed lines are fittings to the smoothed data and the error bars are smaller than the symbols (squares and circles).}
\end{figure}

In order to show the significance of the configurational
disorder on the vibrational properties of the alloy, we depict
in Fig.\ref{S_fr_conf} the vibrational entropy of the alloy in
the fully relaxed case as a function of temperature in
comparison with the total entropy of the alloy in the fully
relaxed case and in the perfectly ordered $L1_2$ phase. In the
fully relaxed case the vibrational entropy is obtained by
subtracting the configurational entropy from the total entropy.
The configurational entropy is, in turn, obtained as a function
of temperature from its mapping with the measured short-range
parameter. The entropy of the $L1_2$ phase is purely
vibrational. Therefore, the difference between the fully
relaxed vibrational entropy and the $L1_2$ vibrational entropy
is only due to configurational disorder. We see that, as the
temperature increases, the vibrational entropy due to disorder
increases steadily and has a finite jump at the OD transition.

The total entropy difference and the configurational entropy
difference at the OD transition in the fully relaxed case are
$\Delta S_{fr}^{tot}=0.35\pm0.02~\rm{k_B/atom}$ and $\Delta
S_{fr}^{conf}=0.27\pm0.01~\rm{k_B/atom}$, respectively. So the
entropy difference excluding the configurational contribution
is $\Delta S_{fr}^{vib}~=~0.08 \pm0.02~\rm{k_B/atom}$. This
result shows that the vibrational entropy difference at the OD
temperature is about 23$\%$ of the total entropy difference
when the volume is allowed to relax. Furthermore, this
vibrational entropy increase is accompanied by a volume
increase of 1.2$\%$. This result, together with the essentially
zero vibrational entropy difference found in the partially
relaxed case, indicates that the volume mechanism is the
responsible for the vibrational entropy increase in $\rm
Ni_3Al$. The increase in volume upon disorder is consistent
with all experimental
\cite{CaClMaMoRo1993,GiNeCa1992,ZhZwBa1995,ZhZwBaref} and
theoretical
work\cite{AnMeMi2007,RaAgBaAnFuHo1998,AlMoFoAsFoJo1998,WaCeWa1998,AnMeMi2007_exp}.
The result that the vibrational entropy difference increases at
the OD transition is consistent with all the experimental
\cite{AnOkFu1993,FuAnNaNiSp1995} and most of the theoretical
work,
\cite{Ac1994,AnMeMi2007,AlMoFoAsFoJo1997,RaAgBaAnFuHo1998,AlMoFoAsFoJo1998}
which have observed a positive vibrational entropy difference
between the totally disordered (metastable) and the totally
ordered phases. The OD temperature in the fully relaxed case is
approximately 30$\%$ lower than the OD temperature calculated
when only the configurational degrees of freedom are
considered. Ozoli\c{n}\v{s}, Wolverton, and Zunger
\cite{OzWoZu1998} proposed a relation between the OD transition
temperature calculated considering only the configurational
degrees of freedom and the transition temperature determined
including also vibrations,
\begin{equation}
T_{od}^{conf+vib}\approx T_{od}^{conf}\left(1+ \frac{\Delta S_{fr}^{vib}}{\Delta S_{conf}^{conf}}\right)^{-1}.
\label{Tconf+vib}
\end{equation}
Using the results from our calculations as input for Eq.
(\ref{Tconf+vib}), namely, $T_{od}^{conf}=1925~\rm K$, $\Delta
S_{conf}^{conf}=0.18~\rm {k_{B}/atom}$, and $\Delta
S_{fr}^{vib}=0.08~\rm {k_{B}/atom}$, we find
$T_{od}^{conf+vib}=1333~\rm K$, i.e., the inclusion of
vibrations lowers the OD transition temperature by 31$\%$, with
respect to the purely configurational transition temperature,
which agrees with the lowering of 30$\%$ we have determined in
the fully relaxed case.

\begin{figure}[htbp]
  \includegraphics[scale=0.55]{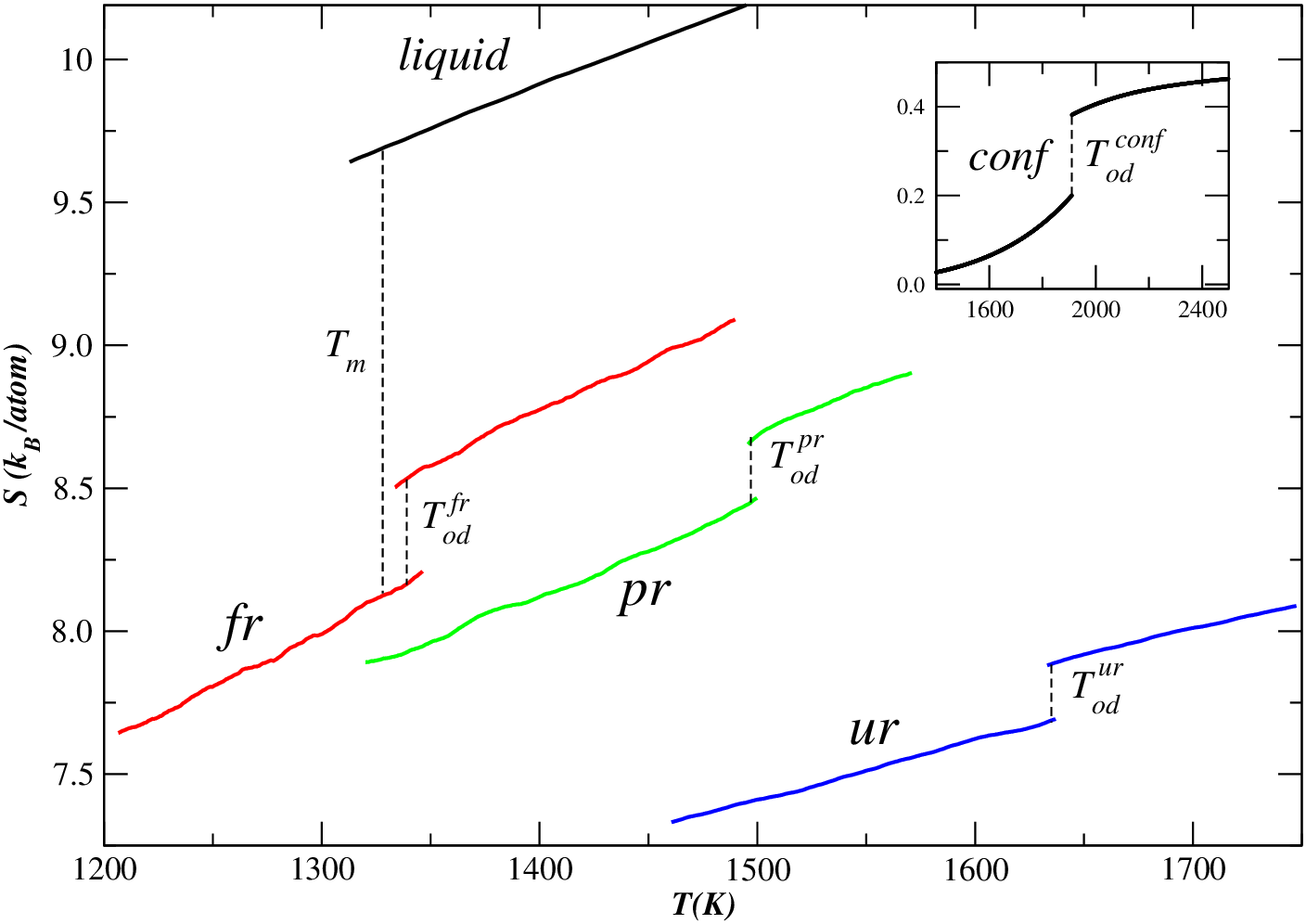}
  \caption{\label{S_off} (Color online) Behavior of the total entropy as a function of
temperature for the liquid phase and all cases studied. The inset shows the configurational case.
The curves are the smoothed data from averages over 10 samples.}
\end{figure}

Finally, in order to compare the changes in entropy for all
cases studied (and also for the liquid phase) we show in Fig.
\ref{S_off} the entropy as a function of the temperature.

\section{SUMMARY}
In this work we employ to the greatest possible advantage the
RS and the Monte Carlo techniques, providing a methodology to
evaluate both the configurational and vibrational free energies
as functions of temperature for n-component substitutional
alloys. This methodology is also used to quantify the
contributions of the vibrational mechanisms by setting
appropriate constraints in the dynamics and calculating the
entropy difference between the relaxed and unrelaxed system. By
consecutively relaxing the configurational and structural
constraints, we are able to quantify the configurational
entropy as well as the vibrational entropy due to the bond
proportion, size mismatch, and volume mechanisms. We applied
this methodology to study the vibrational entropy difference at
the thermodynamical OD transition of the $\rm Ni_3Al$ alloy
obtaining the following results. When the atoms are allowed to
both interchange their positions and vibrate around their ideal
fcc lattice positions, the vibrational entropy difference is
$\Delta S_{ur}^{vib}=-0.07\pm0.02~\rm{k_B/atom}$. This
indicates that bond proportion mechanism decreases the overall
vibration of atoms upon transition. When the atoms can
interchange their positions and are allowed to vibrate around
their equilibrium positions, the vibrational entropy difference
is essentially zero. This indicates that the size mismatch,
coupled to the bond proportion mechanism, does not change the
vibrational entropy upon transition. When the constraints on
atomic interchanging, atomic relaxations, and bulk volume are
relaxed, the vibrational entropy difference at the OD
transition is $\Delta S_{fr}^{vib}=0.08\pm0.02 \;
\rm{k_B/atom}$, which is substantial when compared with the
configurational entropy difference of $\Delta
S_{fr}^{conf}=0.27\pm0.01 \; \rm{k_B/atom}$. This indicates
that the effect of volume relaxation is the source of the
increasing in the overall vibrational entropy upon disorder.
Moreover, the volume increase at the OD transition is of
1.2$\%$. A particular relevant result is that the OD transition
temperature calculated when all constraints are allowed to
relax is approximately 30$\%$ less than that calculated when
only the configurational degrees of freedom are considered.
This result corroborates the importance of the vibrational
degree of freedom in the determination of precise OD phase
diagrams. Finally, as this methodology is not restricted to a
particular crystal structure and stoichiometry, it can be
applied to any n-component substitutional alloy to evaluate the
configurational and vibrational entropies as function of the
temperature and quantify the importance of the corresponding
vibrational mechanisms.

\section*{ACKNOWLEDGEMENTS}
This work was supported by grants from Brazilian Agencies:
CNPq, FAPESP, and FAEPEX/UNICAMP. Part of the computations were
carried out at CENAPAD-SP.


\newpage

\newpage
\newpage

\newpage

\newpage

\newpage

\newpage

\end{document}